\documentclass[a4paper, 12pt]{article}

\usepackage{amsmath,amssymb,amsthm}
\usepackage{mathtools}
\usepackage[utf8]{inputenc}
\usepackage{enumerate}
\usepackage[titletoc]{appendix}
\usepackage{graphicx}
\usepackage{subcaption}
\usepackage[section]{placeins}
\usepackage{setspace}

\usepackage{authblk}

\title{Multiple propagation paths enhance locating the source of diffusion in complex networks}
\author[1]{Ł. G. Gajewski \thanks{lukaszgajewski@tuta.io}}
\author[1]{K. Suchecki}
\author[1,2]{J.A. Hołyst}

\affil[1]{Center of Excellence for Complex Systems Research, Faculty of Physics, Warsaw University of Technology,
Koszykowa 75, 00662 Warsaw, Poland}
\affil[2]{ITMO University, 49 Kronverkskiy av., 197101 Saint Petersburg, Russia}

\date{}

\newcommand{\Tau}{\mathcal{T}}

\begin{document}
\onehalfspacing
\maketitle

\captionsetup[subfigure]{justification=centering}

\begin{abstract}
We investigate the problem of locating the source of diffusion in complex networks without complete knowledge of nodes' states.
Some currently known methods assume the information travels via a single, shortest path, which by assumption is the fastest way.
We show that such a method leads to the overestimation of propagation time for synthetic and real networks, where multiple shortest paths as well as longer paths between vertices exist.
We propose a new method of source estimation based on maximum likelihood principle, that takes into account existence multiple shortest paths.
It shows up to $1.6$ times higher accuracy in synthetic and real networks.

\end{abstract}

\section{Introduction}
Understanding how information propagates in a system is an important field of study in complex networks. The information can be of a various nature - e.g. it could be a virus \cite{sir} or a tweet \cite{tweet}. The main branch of studying epidemics and diffusion is about modelling the propagation itself \cite{forward1,forward2,forward3}. We consider a reverse problem - locating the source, i.e. "the patient zero". That reverse problem has already been approached before. Shah and Zaman introduced a measure of \textit{rumour centrality} defined as the number of distinct ways a rumour can spread from a given node, thus providing with a maximum likelihood estimator of a rumour's source. That method, however, requires the full knowledge of both topology of a network and states of all vertices \cite{reverse}. Pinto, Thiran, Vetterli (PTV), on the other hand, while still requiring full knowledge of network's topology, provide a solution with information about the states of only some of the nodes needed \cite{pinto}. An alternative approach to PTV that retains the limited knowledge assumptions has been introduced by Shen et al. which provides the source estimator via minimizing the variance of a time vector of a backwards spread signal \cite{time-reversal}.
More recent research shows that there are possible solutions for locating a source in temporal networks \cite{temporal} and that an efficient placement of observers is not a trivial task because different common strategies do not significantly differ in resulting accuracy \cite{eff-obs}. The aforementioned PTV method has also been recently improved by Paluch et al. in both accuracy (for scale-free networks) and computation time by limiting observers to closest ones instead of using all, and using gradient to select nodes that have likelihood calculated at all \cite{gmla}.

\section{Set up}
Consider an arbitrary graph that can represent cities connected via highways or friends on Facebook etc. One of the nodes sends some information or a signal to its neighbours.
Then those newly ``infected'' vertices keep passing the message on until all nodes receive it.
We can either assume that the signal is transmitted to node's neighbours with a certain probability each time step (SI model) or that it is always transmitted with a random delay from an arbitrary distribution (Gaussian in our study).
We assume that we know the topology of the network as well as the distribution from which the delays on the links are sampled.
Additionally some subset of all vertices provide us information about at what time they received the signal (we shall call those nodes \emph{observers}).
Our goal is to locate the source of the signal (i.e. the node that generated it).

\begin{figure}[!htb]
\centering
		\includegraphics[width=0.8\textwidth]{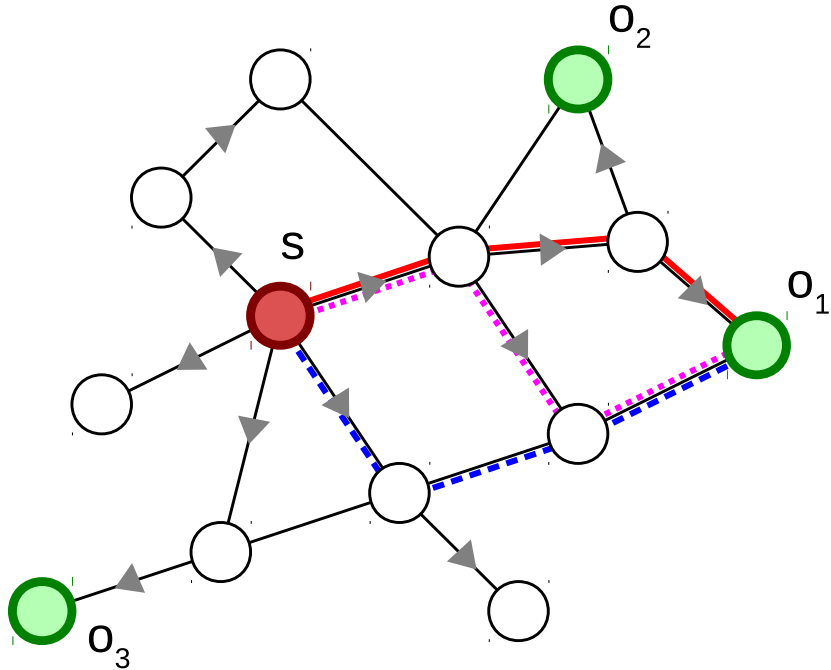}
		\caption{A schematic illustration of the problem. Vertex $S$ is the source we are looking for. Nodes $O_i$ are our observers. One can clearly see that there are three shortest topological paths between the source and vertex $O_1$. Moreover those paths are correlated, i.e. there are edges that are common for several shortest paths between $S$ and $O_1$. What is more, since delays on edges are random variables a shorter path is not always faster, e.g. in the illustration the shortest path between $S$ and $O_2$ is not the fastest one and so the signal travels via a topologically longer route.}
		\label{fig:pretty_fig}
\end{figure}

\begin{figure}[!htb]
\centering
		\includegraphics[width=1.0\textwidth]{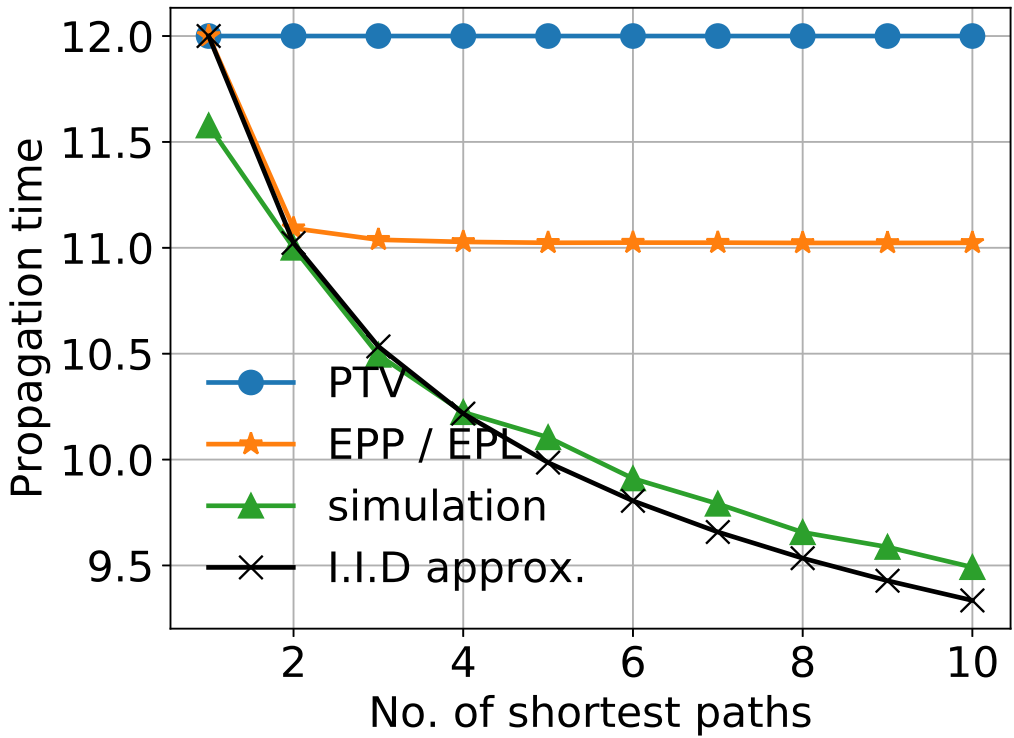}
		\caption{Signal's propagation time between two nodes in a network as a function of number of shortest paths connecting those nodes. Simulations were conducted on Barabási–Albert graph with 100 nodes, with propagation ratio $\frac{\mu}{\sigma} = 4$, two nodes were chosen so that their shortest paths length is set to three edges. Line labelled \textit{simulation} is the real, i.e. experimental time whereas \textit{PTV} is time assumed in that method (shortest path length times $\mu$), \textit{EPP / EPL} is result of taking into account up to two shortest paths (that is what we use for our modifications) see Eq.~(\ref{eq:mean_min}) and \textit{I.I.D} is what we receive when taking all shortest paths into consideration yet assuming their uncorrelated (do not have common edges).}
		\label{fig:ptv_vs_real}
\end{figure}

PTV algorithm assumes that information spreads through shortest paths and while it is intuitively satisfying it is also just an approximation.
In the actual process, there may exist multiple paths of propagation, not only a single one (see Figure~\ref{fig:pretty_fig}).
Assuming that on each edge we have an unknown delay from a known probability distribution and that this distribution is Gaussian (as assumed by Pinto et al.) we can compare PTV's expected time of information's traversal between two nodes to the simulated one. Such comparison is shown at Figure~\ref{fig:ptv_vs_real}. 
There are two reasons for the discrepancy between PTV and the simulation.
First, there can be more than one shortest path (and usually this is the case in non-tree networks).
Since the information travels via \emph{fastest} path, every additional parallel path gives the process another chance for the information to arrive quickly.
Instead of distribution of single path traversal time, we have distribution of minimum time from among several different path times.
Even if mean for each path is the same, the minimum among them has a different distribution than a single path, with a smaller mean value.
Second reason is the existence of independent longer paths, that may end up faster than all shortest paths by chance.
Thus existence of these additional chances for quick traversal (even if much less probable than shortest paths) still decrease mean time.
If the variance of the times compared to mean is small, contribution of longer paths is negligible, but increases as variance becomes larger.
We have focused on the first of the two reasons of discrepancy and developed a new method based on maximum likelihood, centred around existence of multiple shortest paths.
Similar to PTV, we assume normal distributions of delays on edges, and assume that resulting distribution of arrival at different observers is still multivariate normal distribution, or at least it can be approximated as such.
We have therefore to find (i) the mean times of traversal between source and all observers, (ii) the covariance matrix of these times.

Let us assume we have a general graph and each edge has a delay sampled from a known Gaussian distribution $N(\mu, \sigma^2)$.
Particular delays shall remain unknown to the source locating algorithm.
Some randomly chosen subset of nodes (observers) register time of reception as sum of weights along the fastest path between source of information and node in question.
The maximum likelihood for multivariate normal distribution of times would be:
\begin{equation}
\hat{s} = \operatorname*{argmax}_{s \in G} \frac{1}{\sqrt{\Tau}^{n} \sqrt{\left|\pmb{\Sigma}_s\right|}} \exp \left( -\tfrac{1}{2} (\pmb{d}-\pmb{\mu}_s)^T \pmb{\Sigma}_s^{-1} (\pmb{d}-\pmb{\mu}_s) \right)
\label{eq:fullest}
\end{equation}
where $\pmb{d}$ is observed delay vector (each element corresponds to a time at which given observer received the information relatively to the reference observer), $\Tau = 2\pi$, $\pmb{\mu}_s$ is a deterministic delay vector (each element is an expected time of transmission relative to reference observer) if $s$ is the source, $\pmb{\Sigma}_s$ is a covariance matrix assuming source $s$, $\hat{s}$ is our estimated source and $G$ is the whole graph.
PTV assumes information always spreads along single set of shortest paths, that form a breadth-first search tree ($BFS$) rooted at a candidate node $s$.
Deterministic delays $\pmb{\mu}_s$ in such case are equal to $\mu * \pmb{L}_s$, where $\pmb{L}_s$ are lengths of the shortest (and the only existing) paths between source $s$ and observers.
Covariance matrix elements are intersections of shortest path between two given observers to the reference observer (and simply lengths of paths on the diagonal) multiplied by $\sigma^2$.
We shall reject the $BFS$ approach and deal with whole sets of shortest paths, evaluating $\pmb{\mu}_s$ and $\pmb{\Sigma}$ differently.
 
\section{Independent paths}
Let all paths between two nodes be represented by Gaussian random variables $X_1, X_2, \dots, X_n$ with known joint probability distribution
\begin{equation}
f(\pmb{X}) = \frac{\exp \left(-\frac{1}{2}\left(\pmb{X}-\pmb{\mu}\right)^T\,\pmb{\Sigma}^{-1}\left(\pmb{X}-\pmb{\mu}\right)\right)}{\sqrt{\Tau^n\left|\pmb{\Sigma}\right|}}
\end{equation}
Where  $\pmb{X} = [X_1, X_2, \dots, X_n]$, $\pmb{\mu}$ is mean values vector and $\pmb{\Sigma}$ is a covariance matrix. We would love to have an expression for an expected value of propagation time - $\mathbf{E}[X_{min}] = \mathbf{E}[\min(\pmb{X})] = \mathbf{E}[\min(X_1, X_2, \dots, X_n)]$. The probability density function is given by
\begin{equation}
\phi(X_{min}) = -\frac{d}{dX_{min}}P(X_1, X_2,\dots,X_n > X_{min}) = 
\label{eq:pdf_min}
\end{equation}
\begin{equation*}
= -\frac{d}{dX_{min}} \int\limits_{X_{min}}^\infty \dots \int\limits_{X_{min}}^\infty f(\pmb{X}) dX_1dX_2\dots\,dX_n
\end{equation*}
then the expected value $X_{min}$
\begin{equation}
\mathbf{E}[X_{min}] = \int\limits_{-\infty}^\infty x\phi(x)dx
\end{equation}
One can clearly see this is far from a trivial task. We are forced to make some simplifications.
One possible approach is to assume there are no correlations amongst $X_i$, i.e. we deal with independently, identically distributed variables (I.I.D): 
\begin{equation}
\Phi_{X_{min}}(x) = P(X_{min} \le x) =  1 - P(\min(X_1, X_2, \dots, X_n) > x)
\end{equation}
where $\Phi_{X_{min}}(x)$ is the cumulative distribution function of $X_{min}$, also
\begin{equation}
\min(X_1, X_2, \dots, X_n) > x \iff \forall_{i=1,2,\dots,n}\,X_i > x
\end{equation}
then
\begin{equation}
\Phi_{X_{min}}(x) = 1 - P(X_1 > x)P(X_2 > x)\dots P(X_n > x) = 1 - P(X_1 > x)^n =
\end{equation}
\begin{equation*}
= 1 - \left[1 - P(X_1 \le x) \right]^n = 1 - \left[1 - \Phi_{X_1}(x)\right]^n
\end{equation*}
where $\Phi_{X_1}$ is given by
\begin{equation}
\Phi_{X_1}(x) = \frac{1}{2}\left(1 + \mathrm{erf}\left(\frac{x-\mu}{\sqrt{2}\sigma}\right)\right)
\end{equation}
The probability density function of $X_{min}$ is then given by
\begin{equation}
\phi_{X_{min}}(x) = \frac{d\Phi_{X_{min}}(x)}{dx}
\end{equation}
After some working out we conclude that
\begin{equation}
\mathbf{E}[X_{min}] = \int\limits_{-\infty}^{\infty}x\phi_{X_{min}}(x)dx = 
\label{eq:iid_xmin_gauss}
\end{equation}
\begin{equation*}
= \int\limits_{-\infty}^{\infty} 
\frac{x n \exp\left(-\frac{(x-\mu)^2}{2\sigma^2}\right)}{\sqrt{\Tau}\sigma}
\left[1 - \frac{1}{2}\left(1 + \mathrm{erf}\left(\frac{x - \mu}{\sqrt{2}\sigma}\right) \right) \right]^{n-1} dx
\end{equation*}

Alternatively we can also take into account that negative delays are non-existent in vast majority of real world scenarios (i.e. $x \ge 0$) and then we have
\begin{equation}
\Phi_{X_1}(x) = \frac{\mathrm{erf}\left(\frac{x-\mu}{\sqrt{2}\sigma}\right)+\mathrm{erf}\left(\frac{\mu}{\sqrt{2}\sigma}\right)}{1+\mathrm{erf}\left(\frac{\mu}{\sqrt{2}\sigma}\right)}
\end{equation}
\begin{equation}
\mathbf{E}[X_{min}]  = \int\limits_{-\infty}^{\infty}
\frac{x n \exp\left(-\frac{(x-\mu)^2}{2\sigma^2}\right)}{\sqrt{\Tau}\sigma}
\left[1 - \frac{\mathrm{erf}\left(\frac{x-\mu}{\sqrt{2}\sigma}\right)+\mathrm{erf}\left(\frac{\mu}{\sqrt{2}\sigma}\right)}{1+\mathrm{erf}\left(\frac{\mu}{\sqrt{2}\sigma}\right)} \right]^{n-1} dx
\label{eq:iid_xmin}
\end{equation}
Ignoring that, however, can be justified if \emph{propagation ratio} - $\frac{\mu}{\sigma}$ is sufficiently large (due to the nature of \emph{error function}).

Comparison of the above approximation (\ref{eq:iid_xmin_gauss}) with simulation, PTV and alternative approach chosen for our modifications is shown at Figure~\ref{fig:ptv_vs_real}. Results are for propagation ratio $\frac{\mu}{\sigma} = 4$ and one could say they are promising, however, for large values of $\sigma$ or for strongly correlated paths the results are less satisfying. While that provides a vast improvement over original PTV approach the nature of the integral is such that it takes substantially more time to calculate and so its benefits might not necessarily exceed its costs.

\section{Multiple correlated paths}
Since establishing general expression for the expected value of minimum distribution of an ensemble of $n$ shortest paths is a non-trivial task, even with I.I.D approximation described above, we shall settle for a known analytical formula in the case of $n=2$ derived in \cite{min}.

Let $Y = min(X_1, X_2)$ where $X_1, X_2$ are known Gaussian distributions such that $E[X_i] = \mu_i, E[X_i^2]-E[X_i]^2 = \sigma_i$, and $\Phi, \phi$ are respectively the cumulative distribution function and probability d.f. of the standard normal distribution. Then:
\begin{equation}
E[Y] = \mu_1\Phi\big(\frac{\mu_2 - \mu_1}{\theta} \big) + \mu_2\Phi\big(\frac{\mu_1 - \mu_2}{\theta} \big) - \theta\phi\big(\frac{\mu_2 - \mu_1}{\theta} \big)
\label{eq:mean_min}
\end{equation}
\begin{equation}
E[Y^2] = (\sigma_1^2 + \mu_1^2)\Phi\big(\frac{\mu_2 - \mu_1}{\theta} \big) + (\sigma_2^2 + \mu_2^2)\Phi\big(\frac{\mu_1 - \mu_2}{\theta} \big) - (\mu_1 + \mu_2)\theta\phi\big(\frac{\mu_2 - \mu_1}{\theta} \big)
\label{eq:var_of_min}
\end{equation}
\begin{equation}
\theta = \sqrt{\sigma_1^2 + \sigma_2^2 - 2\rho\sigma_1\sigma_2}
\end{equation}
Where $\rho$ is a correlation coefficient between the distributions $X_1, X_2$ that in case of paths we define as number of common edges between the paths $R_1, R_2$ normalized by the length $L$ of the path:
\begin{equation} 
\rho = \frac{R_1 \cap R_2}{L}
\end{equation}
Now evaluate elements of $\pmb{\mu}$ using (\ref{eq:mean_min}) if $n > 1$ else as in PTV. In case there are more than two shortest paths apply (\ref{eq:mean_min}) for two least correlated ones (that is those who have the smallest number of common edges). Result of such approach compared with PTV and "real" times in simulation is presented at Fig.~\ref{fig:ptv_vs_real}. As one can clearly see it is still far from truth yet significantly closer than PTV.

\begin{figure}[!htb]
\centering
		\includegraphics[width=0.8\textwidth]{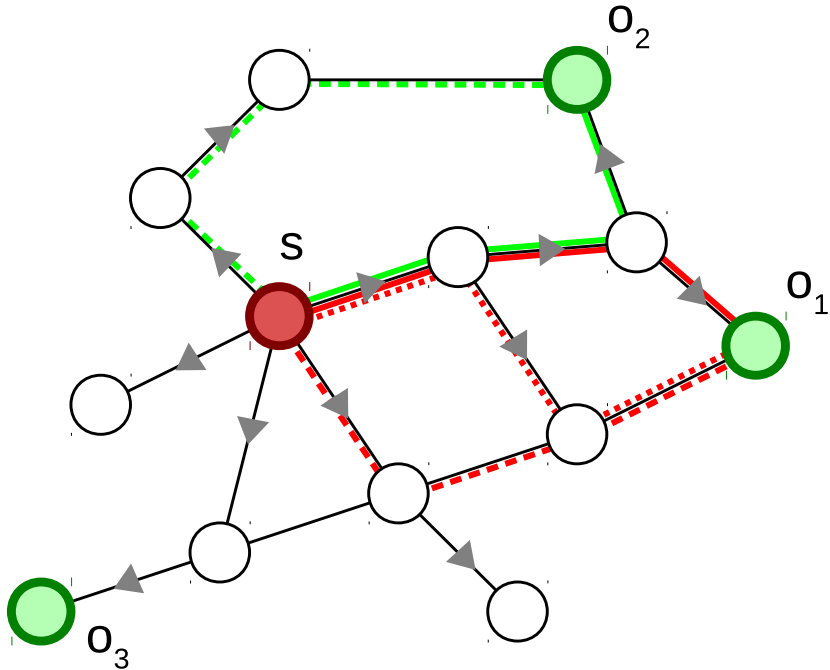}
		\caption{Correlations between paths between supposed source $S$ and observers $O_1$ (red paths) and $O_2$ (green paths). Approach EPP assumes all paths are equally probable and calculates expected overlap between all pairs of red-green paths (out of six pairs, one has overlap 1, one has overlap 2). Approach EPL assumes all links are equally probable to be used by real spreading and calculates overlap between union of all red and union of all green paths (two links), normalized by union of all paths (eleven links).}
		\label{fig:modifications}
\end{figure}

\subsection{Equiprobable paths (EPP)}
Since we have rejected $BFS$ trees we cannot calculate covariance matrices as before. A natural generalization of PTV's formula is to use expected value of covariance between all shortest paths from node $o_{i+1}$ to reference observer $o_1$ and all shortest paths between $o_{j+1}$ and $o_1$ while on diagonal elements instead of $\sigma^2 * L$ we calculate the variance of minimum distribution if $n > 1$ (Fig. \ref{fig:modifications}). I.e.:
Let $\heartsuit$ represent a set of paths each treated as random variable:
\begin{equation}
\heartsuit_{i+1} := R_1(o_1, o_{i+1}), R_2(o_1, o_{i+1}), \dots, R_n(o_1, o_{i+1})
\end{equation}
where $R_i(x, y)$ is an $i-th$ path between vertices $x$ and $y$, then:
\begin{equation}
\Sigma_{i,j} = 
\left\{\begin{array}{ll}
\sigma^2 \sum_{k=1}^{n}\sum_{l=1}^{m}p_{k,l}\cdot |R_{k}(o_1, o_{i+1})\cap R_{l}(o_1, o_{j+1})| & i\neq j \\
\min_{\sigma^2}(\heartsuit_{i+1} ) = E[Y^2] - E[Y]^2 & i = j \\
\end{array}\right.
\label{eq:cov_hearts_gen}
\end{equation}
Where $p_{k,l}$ is a probability that $k-th$ path from the set of paths connecting nodes $o_1, o_{i+1}$ and $l-th$ path from the set connecting the other observer with reference observer (i.e. $o_1, o_{j+1}$) will be the fastest routes, see Fig.~\ref{fig:modifications} for details.
Unfortunately those probabilities are non-trivial to evaluate so we assume all paths to be equally probable (which is true only when there are no intersection between them).
With this assumption, we obtain
\begin{equation}
\Sigma_{i,j} = 
\left\{\begin{array}{ll}
\frac{\sigma^2}{n\cdot m}\sum_{k=1}^{n}\sum_{l=1}^{m} |R_{k}(o_1, o_{i+1})\cap R_{l}(o_1, o_{j+1})| & i\neq j \\
\min_{\sigma^2}(\heartsuit_{i+1}) & i = j \\
\end{array}\right.
\label{eq:cov_hearts}
\end{equation}

\subsection{Equiprobable links (EPL)}
Alternatively one can use a product of Jaccard Index \cite{jaccard} (Fig. \ref{fig:modifications}) and geometric mean of variances of sets of paths, namely:
\begin{equation}
\Sigma_{i,j} = \frac{|\{e_i\}\cap \{e_j\}|}{|\{e_i\}\cup\{e_j\}|}\cdot\sqrt{\min_{\sigma^2}(\heartsuit_{i+1})\cdot\min_{\sigma^2}(\heartsuit_{j+1})}
\label{eq:cov_spades}
\end{equation}
where $\{e_i\}$ is a set of edges of all shortest paths connecting node $i$ with the reference observer. Intuitively one could say that using \textit{intersection over union} of sets of edges represents treating each edge as equally probable (instead of treating each \textit{path} equally probable like in previous approach). The choice of geometric mean is arbitrary and arithmetic mean produces similar results.

\section{Results}
\begin{figure}
\centering
	\begin{subfigure}[!htb]{0.5\textwidth}
		\includegraphics[width=\textwidth]{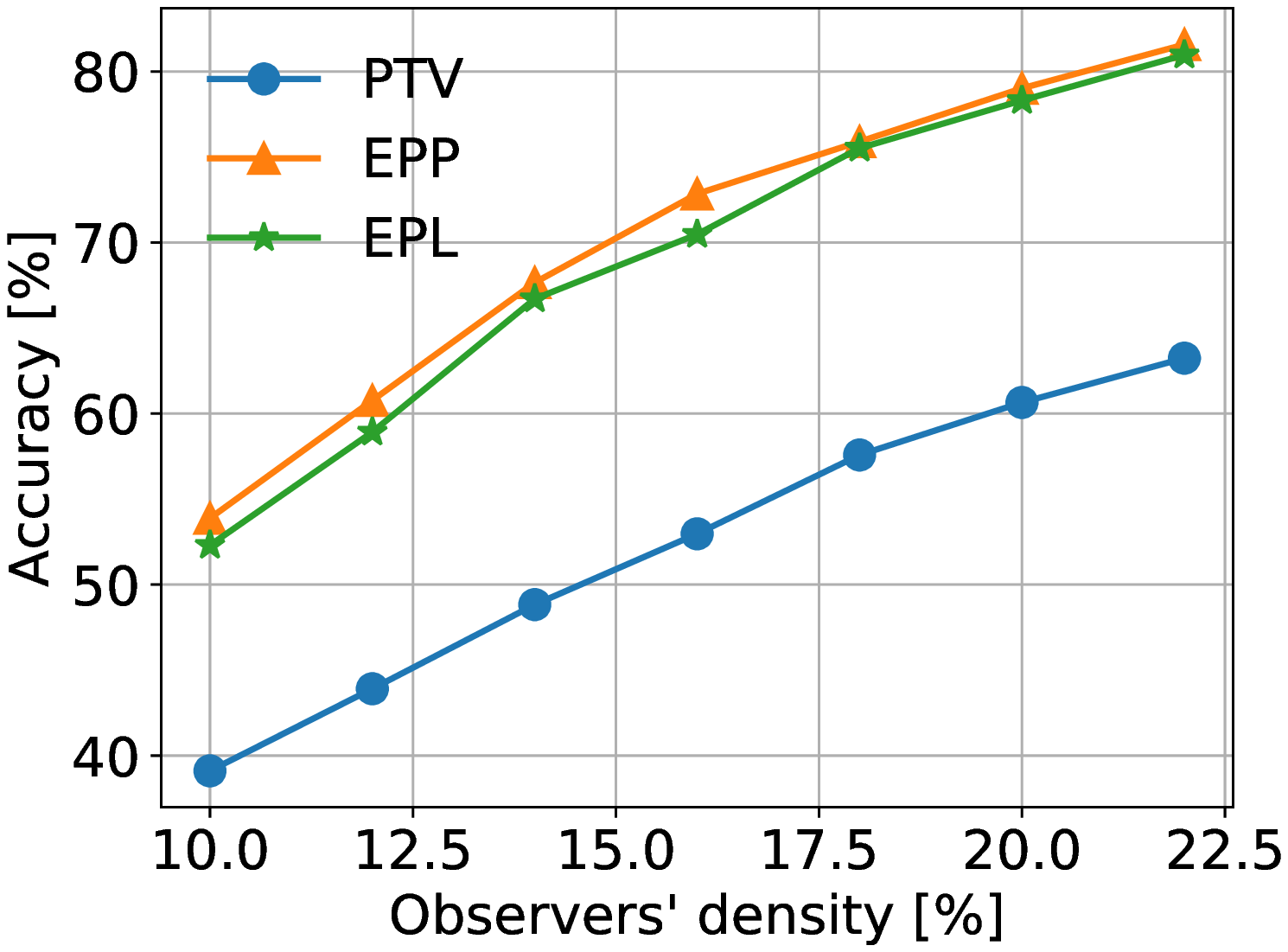}
		\caption{}
		\label{fig:ba_3_1}
	\end{subfigure}~
	\begin{subfigure}[!htb]{0.5\textwidth}
		\includegraphics[width=\textwidth]{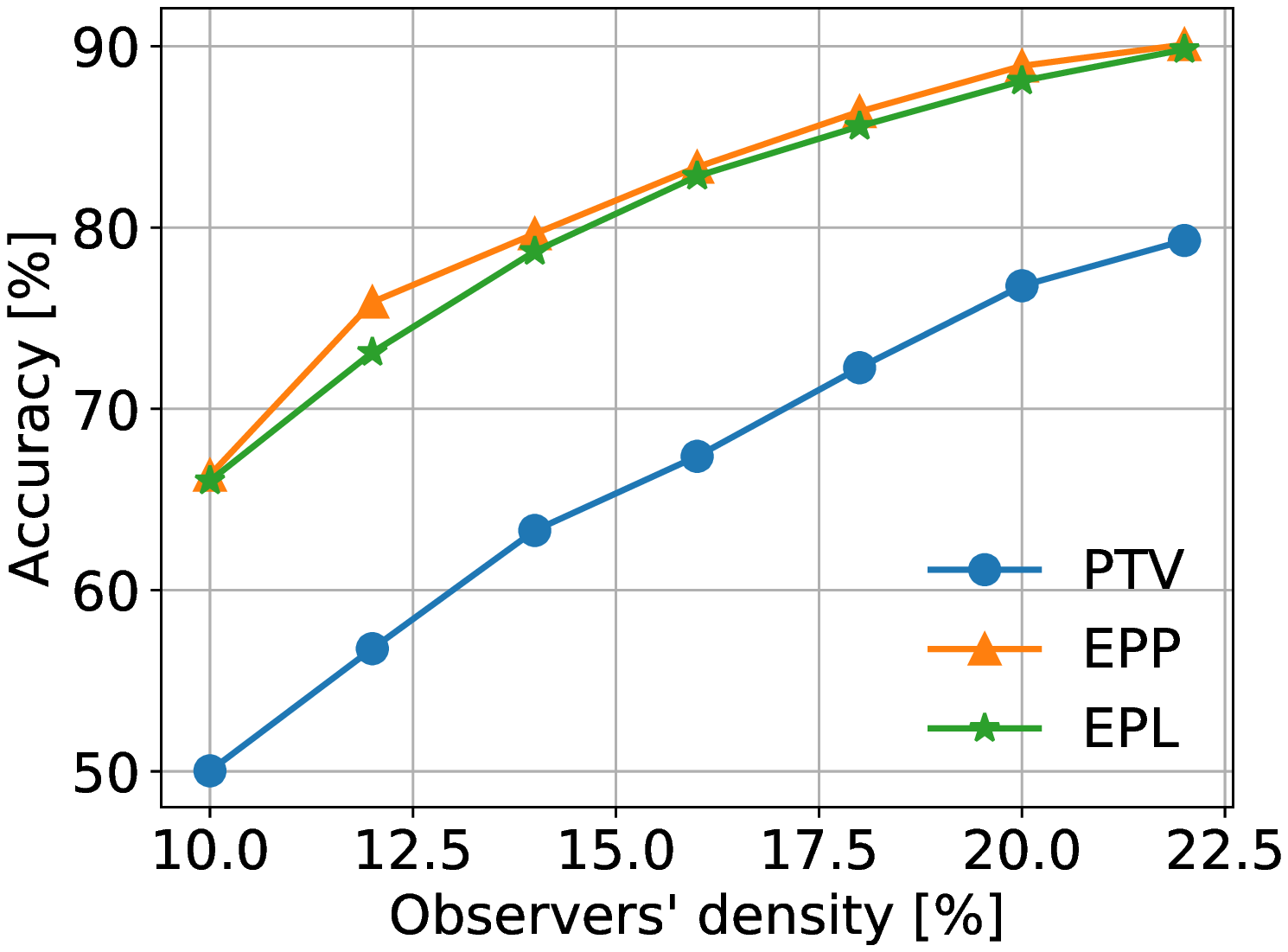}
		\caption{}
		\label{fig:er_6_1}
        \end{subfigure}
        \caption{Improved accuracy of source detection measured as a function of observers' density for original PTV method and and our modifications. Simulations where conducted on (a) Barabási–Albert (BA) and (b) Erdős–Rényi (ER) graph. Both with mean node degree $\langle k \rangle=6$ and network size $N=100$. We have conducted 100 simulations for each point on those plots.}
        \label{fig:scores}
\end{figure}

Presented method has been tested on synthetic graphs and real network.
We used two synthetic graphs: Barabási–Albert (BA) and Erdős–Rényi (ER) with size $N=100$ and average node degree $\langle k \rangle=6$ in each case. Since estimator in both methods can produce two or more nodes with the same maximum score the success of algorithm is registered when the actual source is within the list of those candidates with maximum likelihood. Results are shown at Figure~\ref{fig:scores}. One can clearly see a vast improvement in accuracy of the method using our proposed adjustments with the difference between EPP and EPL being barely noticeable yet still consistently in favour of approach EPP.

\begin{figure}[!htb]
    \centering
        \includegraphics[width=\textwidth]{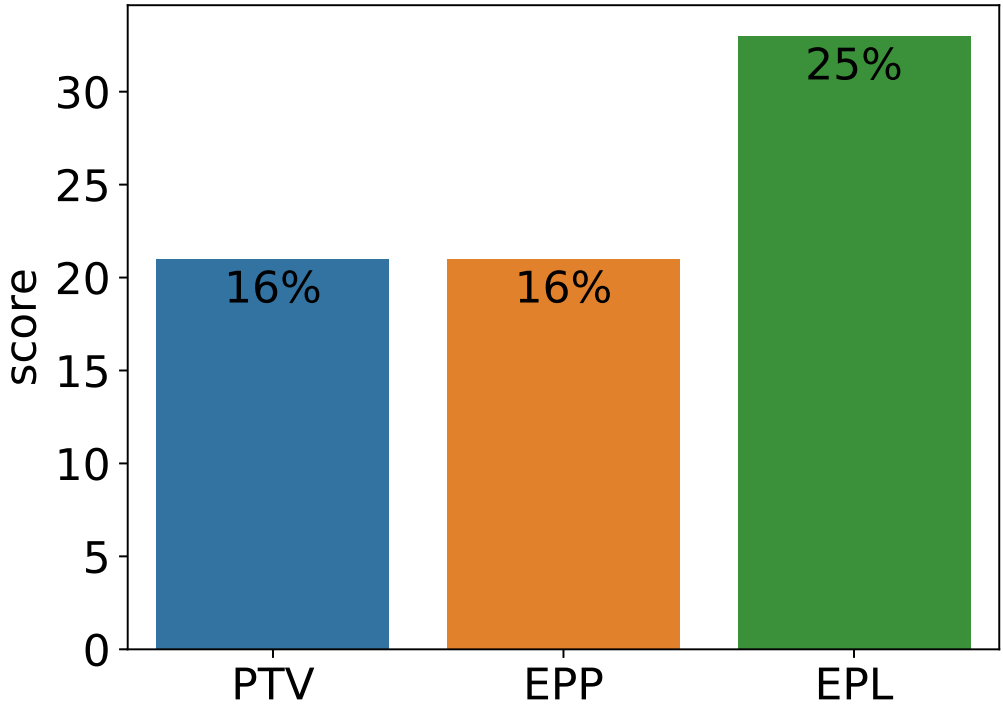}
        \caption{Improved accuracy on real life network. The simulations were conducted on the ego-facebook graph with original PTV method and proposed modifications - EPP and EPL. While PTV and EPP have successfully detected the epidemic source 16\% (21/133) of the time the EPL approach has scored 25\% (33/133).}
        \label{fig:facebook_score}
\end{figure}

We have also conducted tests on a real network: ego-facebook with $N=4039$ \cite{egoface}.
After 133 simulations at observers' density $d=10\%$ EPL successfully located the source $33$ ($25\%$) times while $PTV$ and EPP both scored $21$ ($16\%$)(Fig~\ref{fig:facebook_score}).
In this case the EPP modification performed worse than EPL, on the contrary to the case of synthetic networks.
The reason for that maybe high clustering coefficient of the real network while ER and BA models are known to not reproduce that characteristic.
A high clustering coefficient naturally leads to more correlations between paths and those make our approximation of equally probable paths in approach EPP not sustainable.

\begin{figure}
\centering
	\begin{subfigure}[!htb]{0.3\textwidth}
		\includegraphics[width=\textwidth]{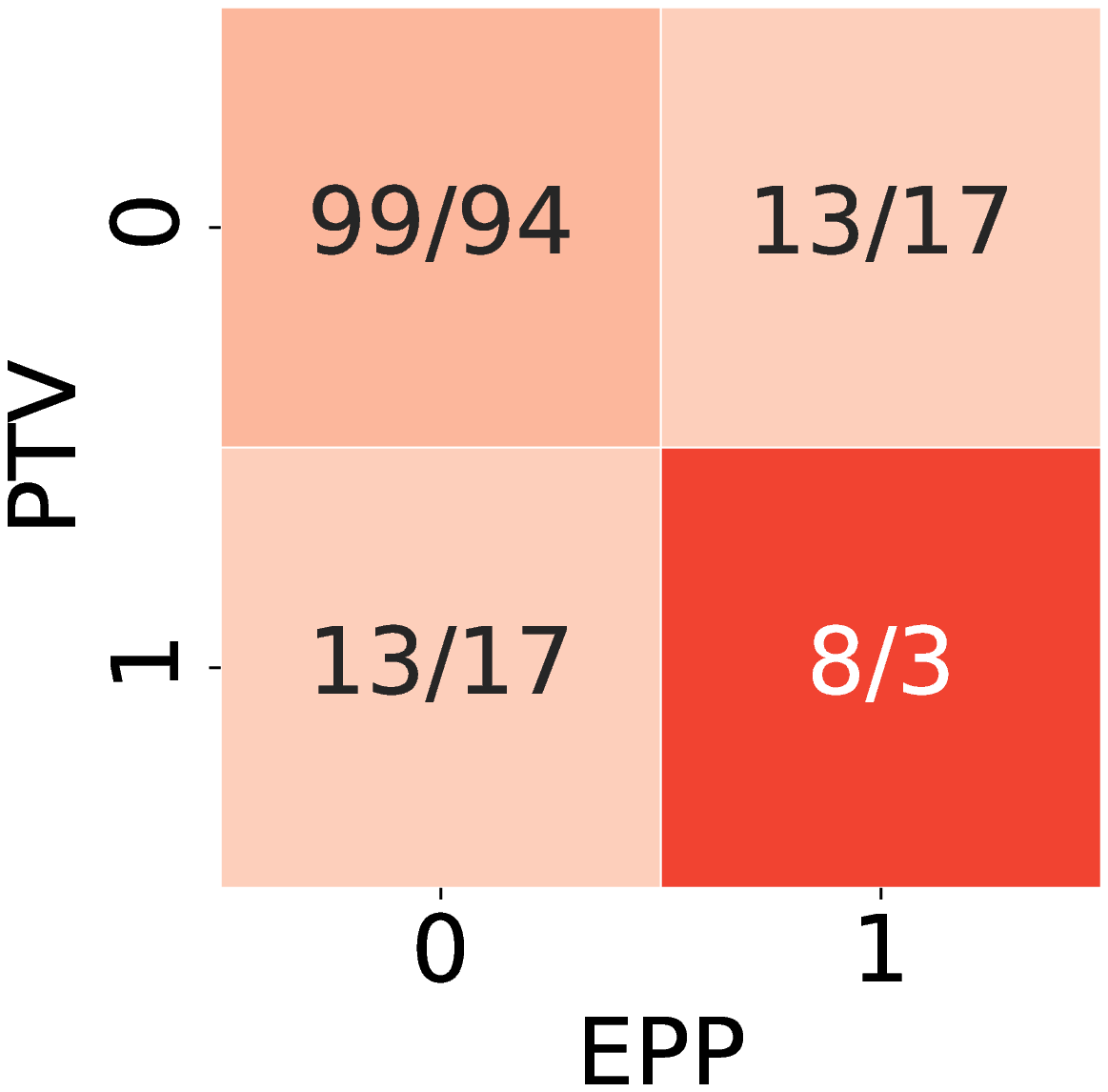}
		\caption{}
		\label{fig:ba_3_2}
	\end{subfigure}~
	\begin{subfigure}[!htb]{0.3\textwidth}
		\includegraphics[width=\textwidth]{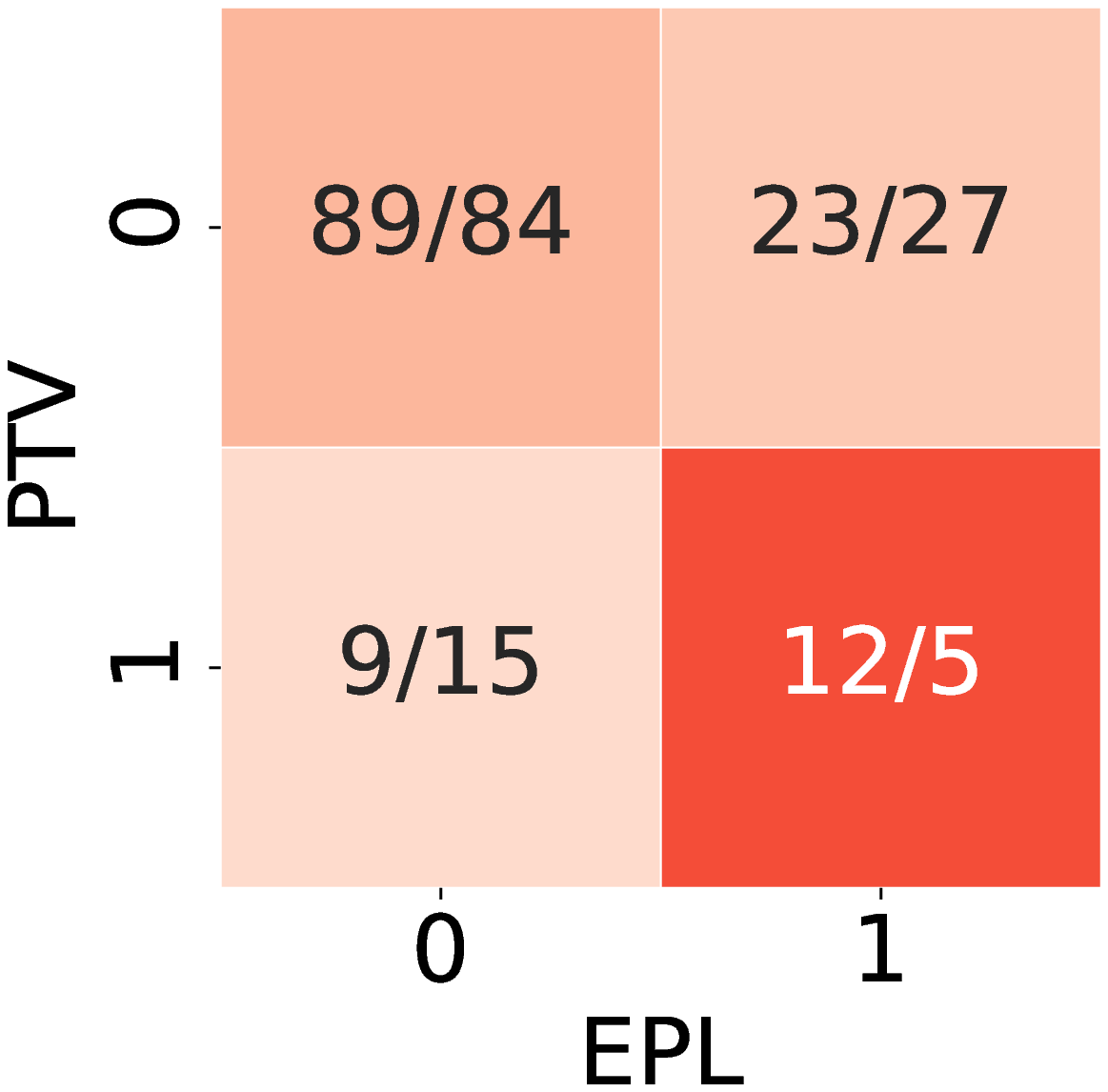}
		\caption{}
		\label{fig:er_6_2}
        \end{subfigure}~
        \begin{subfigure}[!htb]{0.38\textwidth}
		\includegraphics[width=\textwidth]{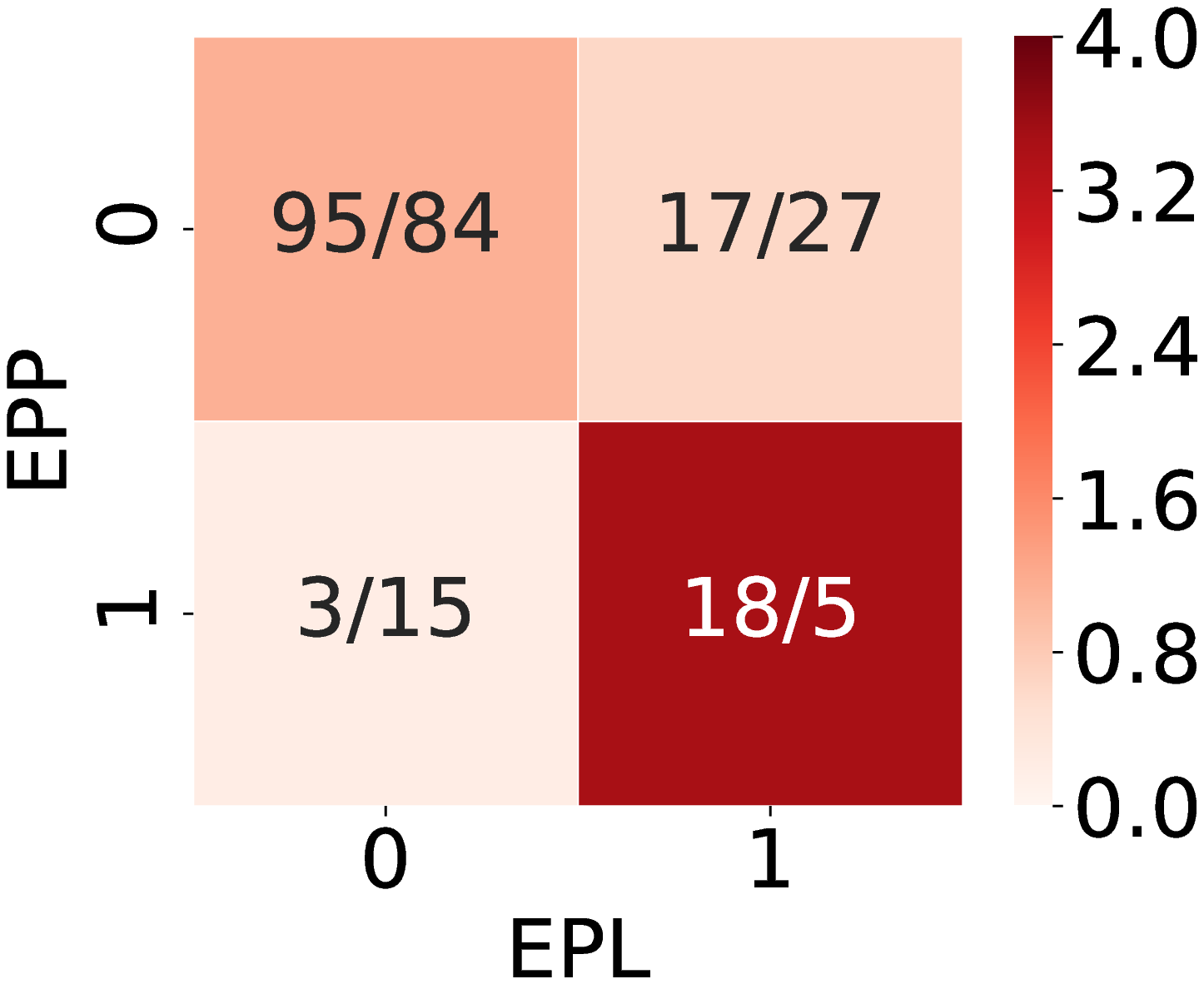}
		\caption{}
		\label{fig:er_6_3}
        \end{subfigure}
     \caption{Contingency matrices. $0$ means the algorithm failed to locate the source, $1$ means it succeeded. The first number of the label is the count of appropriate events while the second is the expected based on methods' performance and that they are independent. Colours are applied according to the fractions of the aforementioned numbers.}
     \label{fig:contingency}
\end{figure}

Contingency matrices for a PTV, EPP, EPL methods are presented at fig~\ref{fig:contingency}. We see that the probability of agreement of any pair of those three methods in successful source location is much higher than in the case they were completely independent. 

\section{Computational complexity}

\begin{figure}[!htb]
    \centering
        \includegraphics[width=\textwidth]{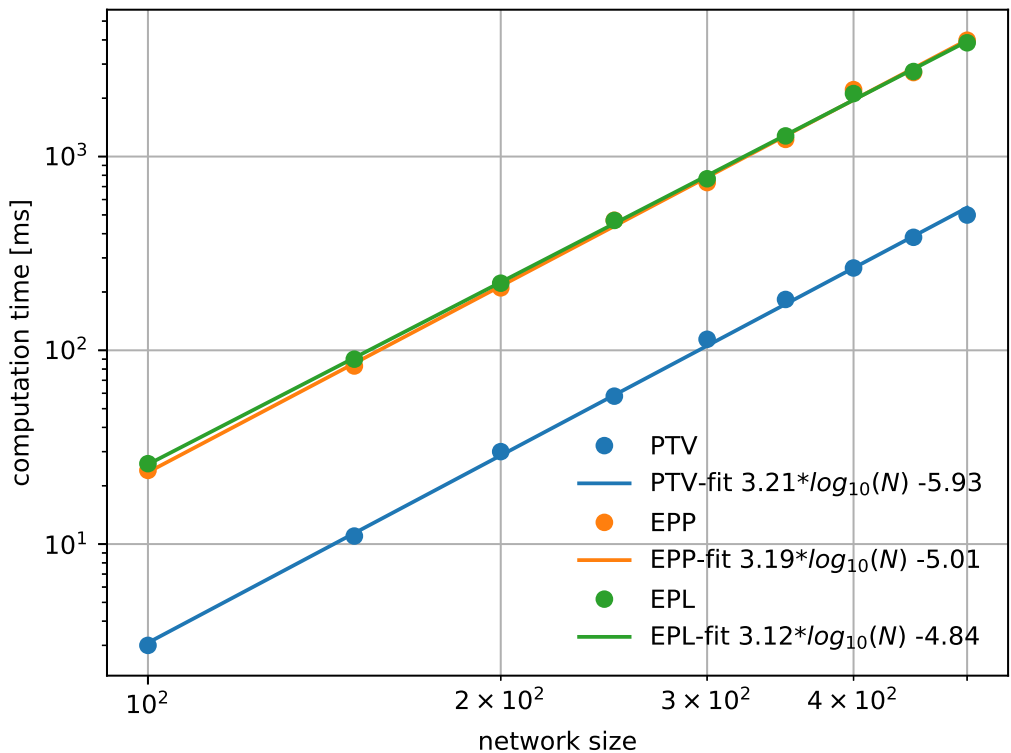}
        \caption{Computation times as a function of network size for original PTV and our modifications. Plot is in logarithmic scale and appropriate linear fitting has been done. Simulations where conducted on Barabási–Albert graph with observers' density $d=0.1$.}
        \label{fig:ba_scaling}
\end{figure}

Pinto et al. report that the computational time in their method scales with network size $N$ as $N^\alpha$ where for arbitrary graphs $\alpha = 3$. In our implementation of PTV we have a reasonably close result of $\alpha = 3.21$. Tests were conducted on a Barabási–Albert graph with observers' density $d=0.1$. Our own modifications seem to slightly improve scalability, i.e. $\alpha_{EPP} = 3.19$, $\alpha_{EPL} = 3.12$. However, both EPP and EPL bring significant initial costs resulting in overall computation time to be significantly higher than original PTV's. See Figure~\ref{fig:ba_scaling} for details. The nature of our modifications also makes theoretical predictions of scalability much harder to acquire. Although in BA example they seem to follow $N^\alpha$ fit, they heavily depend on graph's density, number of shortest paths and correlations between them. As such it might be advisable to do some pre-computing when applying EPP / EPL on real networks. If observers are known beforehand and do not change one would have to build covariance matrix and deterministic delay only once. The only element changing would be the observed delay vector and all computation is reduced to a vector-matrix-vector multiplication.

\section{Conclusions \& Discussion}
We have shown that the approach of breadth-first search ($BFS$) trees presented by Pinto et al. overestimates propagation time of a signal.
We presented an analysis of why that is, namely there can be more than one shortest paths and also other paths can also happen to be faster, and those effects are non-negligible.
We provide alternatives to $BFS$ that take multiple paths into account and while requiring more computation time, they significantly improve accuracy of the source detection.
Improvement in accuracy is mostly prominent for artificial networks (BA, ER), however, when tested on a real network (ego-facebook) there was a visible increase in accuracy ($1.6$ times higher) with approach EPL while EPP was no worse than original PTV method. 
While results of presented methods show a lot of promise there are some obvious paths one can still undertake to improve the accuracy.
Firstly, we would love to have an analytical expression for the minimum distribution of an arbitrary amount of Gaussian distributions and not only two.
Secondly, one could expect that should the probabilities in EPP variant be known (instead assuming they are all equal) the difference in accuracy between EPP and EPL should increase in the favour of the former.
Thirdly, for arbitrary graphs the proposed variants do not scale as trivially with network size as original PTV ($N^{\alpha}$, $\alpha \in [3,4]$) for they heavily depend on number of edges, shortest paths and intersections among them thus an exact expression for that scalability is unknown.
For both methods, appropriate pre-computing can cut on time needed to locate source after observer reports are obtained.
\\
\\
\\
\section*{Acknowledgements}
The work was partially supported as RENOIR Project by the European Union Horizon 2020 research and innovation programme
under the Marie Skłodowska-Curie grant agreement No 691152 and by Ministry of Science and Higher Education (Poland),
grant Nos. W34/H2020/2016, 329025/PnH /2016. and National Science Centre, Poland Grant No. 2015/19/B/ST6/02612. 
J.A.H. was partially supported by the Russian Scientific Foundation, Agreement \#17-71-30029 with co-financing of Bank Saint Petersburg.

\bibliographystyle{plain}
\bibliography{references.bib}

\end{document}